\documentclass[%
reprint,
superscriptaddress,
 amsmath,amssymb,
 aps,
floatfix,
nofootinbib,
]{revtex4-2}

\usepackage{graphicx}
\usepackage{dcolumn}
\usepackage{bm}

\usepackage{xcolor}

\begin{document}

\preprint{APS/123-QED}

\title{Discontinuous Jump Behavior of the Energy Conversion in Wind Energy Systems}

\author{Pyei Phyo Lin}
\email{pyei.phyo.lin@uol.de}
\affiliation{ForWind, Institute of Physics, University of Oldenburg, Oldenburg, Germany}
\author{Matthias W\"achter}
\affiliation{ForWind, Institute of Physics, University of Oldenburg, Oldenburg, Germany}
\author{M. Reza Rahimi Tabar}%
\affiliation{Department of Physics, Sharif University of Technology, Tehran 11155-9161, Iran}
\affiliation{ForWind, Institute of Physics, University of Oldenburg, Oldenburg, Germany}
\author{Joachim Peinke}
 \affiliation{ForWind, Institute of Physics, University of Oldenburg, Oldenburg, Germany}

\date{\today}

\begin{abstract}
The power conversion process of a wind turbine can be characterized by a stochastic differential equation (SDE) of the power output conditioned to certain fixed wind speeds. An analogous approach can also be applied to the mechanical loads on a wind turbine, such as generator torque. The constructed SDE consists of the deterministic and stochastic terms, the latter corresponding to the highly fluctuating behavior of the wind turbine. Here we show how advanced stochastic analysis of the noise contribution can be used to show different operating modes of the conversion process of a wind turbine. The parameters of the SDE, known as Kramers-Moyal (KM) coefficients, are estimated directly from the measurement data. Clear evidence is found that both, continuous diffusion noise and discontinuous jump noise are present. The difference in the noise contributions indicates different operational regions. In particular, we observe that the jump character or discontinuity in power production has a significant contribution in the regions where the control system switches strategies. We find that there is a high increase in jump amplitude near the transition to the rated region, and the switching strategies cannot result in a smooth transition. The proposed analysis provides new insights to the control strategies of the wind turbine.
\end{abstract}

\maketitle



\section{Introduction}
Wind energy is one of the most promising contributions to the global energy transition from fossil fuels to clean and sustainable energy. Europe could install around 105\,GW of new wind energy capacity in the period of 2021--2025 as reported in \cite{Windeurope2021}. However, the complex and intermittent nature of wind makes wind energy production difficult to predict, which is important for a stable energy supply \cite{Muzy2010, Calif2014, Anvari2016b, Schmietendorf2017}. Furthermore this nature of wind may cause premature mechanical fatigue failure \cite{Muecke2011, Waechter2012}. It is known that the commonly used industry standard by the International Electrotechnical Commission \cite{IEC61400} is not describing properly the variability of wind and wind power \cite{Haehne2018, Waechter2012}. In particular, if one investigates time series of the power output of a wind turbine, one can find very rapid power fluctuations, which can become larger than 50\% of the rated power \cite{Milan2013}. Such short time power fluctuations in the range of MW will represent special loads for the drive train and also for the power grid, as these fluctuations seem to some add up in a wind farm instead of being averaged out \cite{Milan2013, Haehne2018}. 

In this contribution, we focus on a statistically advanced description of the power fluctuations of a wind turbine. In recent years, it has been shown that the power conversion process of a wind turbine can be modeled by a stochastic Langevin differential equation of the power output $P$ conditioned to certain fixed wind speeds $u$ \cite{Anahua2008, Gottschall2008, Milan2013}. An analogous approach has also been used to model the mechanical loads on a wind turbine such as generator torque $T$ \cite{Lind2014}. The advantage of this approach is that the model equations (in form of the Langevin equation) can be extracted directly from given data. This model can reproduce the stochastic, turbulent and intermittent nature of wind power \cite{Milan2013}. In the Langevin modeling of conversion dynamics of a wind turbine, the focus was up-to now on the deterministic part of the power time series, while the question remained open how to correctly capture the abrupt large power fluctuations mentioned above.

The Langevin equation describes a diffusion process with continuous trajectory. It consists of the deterministic term and the continuous stochastic term which is modeled by a Wiener process or a Brownian motion. Its two model parameters which are the drift and diffusion coefficients can be estimated directly from the measurement data \cite{Friedrich2011, Rinn2016, Tabar2019}. 
These parameters are also known as Kramers-Moyal (KM) coefficients which are considered up to second order in the Langevin equation. For the continuous process, the coefficients higher than third order are negligible. 

\begin{figure*}[t]
\centering
\includegraphics[width = \textwidth]{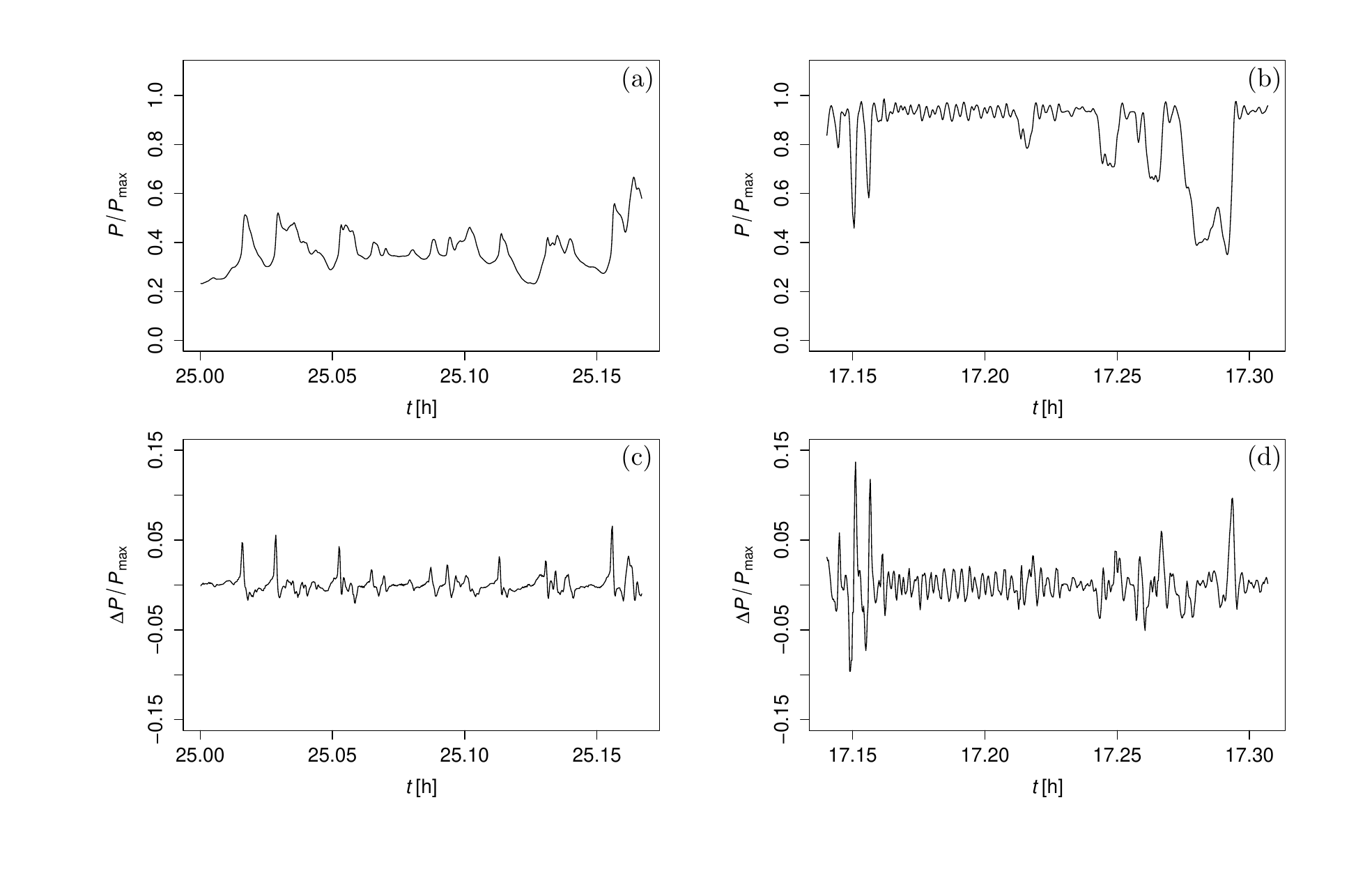}
\caption{Wind power time series, spanning the period of ten minutes. (a) shows the period where the power changes are not very large.(b) shows the period where the power changes are very large, up to about 40\% of the rated power. Increments $\Delta P:=P(t+\tau) - P(t)$ emphasize the fluctuations and are shown for sampling period $\tau=1$s in (c) and (d).}
\label{fig:powerts}
\end{figure*}

Looking at the temporally high-resolved wind power data, one can see portions of time series, which look like a diffusive process (see Fig.~\ref{fig:powerts}~(a), but there are also periods where sudden big jumps of the delivered power become obvious, see Fig.~\ref{fig:powerts}~(b). 

In this contribution, we aim to investigate how far these jumps make it necessary to extend the stochastic description.
If the higher order KM coefficients ($>3$) are not negligible, they would be an indicator of non-continuity in the process \cite{Anvari2016a, Tabar2019}. One possibility to model this behavior is the extension of the Langevin diffusion process to a jump-diffusion process. An additional discontinuous stochastic term for the jump process is introduced for which we assume that it can be modeled by a Poisson process. In this more general stochastic approach two more parameters arise which are the jump rate and the jump amplitude. We show how they can be estimated from the higher order KM coefficients. This analysis aims to give a more realistic stochastic description of the power output of a wind turbine. Relation to control strategies, or the use for improved modeling of the wind energy resource in a power grid, will be discussed in this paper.

Our aim is to show in detail the procedure to estimate the general stochastic jump-diffusion process with Wiener and Poisson noise to achieve an advanced stochastic characterization and modelling of the wind power conversion dynamics of a wind turbine. Here, we analyze the SCADA data from a wind turbine with resolution of $1~\mathrm{Hz}$. The article is organized as follows. At first, we describe the analysed data. Next, the stochastic analysis method is summarized, and it is shown how it is possible to quantify and separate the contributions of diffusion and jump fluctuations. Finally, the results of the data analysis for power output conditioned to wind speed are presented. In addition, we investigate the stochastic relation between generator torque and generator rotational speed.

\section{Stochastic data analysis of wind energy system}

\subsection{Data description}
\label{sec:data}

The measurement data are extracted from a wind turbine of a wind farm. The wind farm is installed onshore over an area covering roughly $4\,{\rm km^2}$ and is surrounded by flat rural terrain with 12 identical variable-speed, pitch-regulated wind turbines. The rated power of each turbine is in the order of $2\,{\rm MW}$. The values were made anonymous to keep the confidentiality of the data. Thus, all the data are normalized with their corresponding maximum for our analysis.

The measured quantities are the net electrical power output, $P$, generated by the wind turbine, the wind speed, $u$, measured on the nacelle by a cup anemometer and the rotational speed or rpm, $\Omega$, of the generator. The torque, $T$, on the generator is calculated from the power and rpm of the generator using the relation
\begin{equation}
    T = \frac{60\,{\rm s}}{2\pi} \frac{P}{\Omega}~.
    \label{eq:torq}
\end{equation}
All measurements were performed at a sampling frequency $f_s = 1\,{\rm Hz}$. The measurement campaign was conducted over a period of eight months, from June 2009 till February 2010. The same data were used also in the study of \cite{Milan2014}. 

\subsection{Power Conversion Process Described by Stochastic Dynamics}
\label{sec:sde}

Assuming the validity of a diffusive process, the power conversion process of a wind turbine can be modelled as a stochastic Langevin equation of the power output $P$ conditioned to certain fixed wind speed $u$ \cite{Anahua2008, Gottschall2008, Milan2013}, 
\begin{equation}
\mathrm{d}P(t, u) = D^{(1)}(P \vert u) \, \mathrm{d}t + \sqrt{D^{(2)}(P \vert u)} \, \mathrm{d}W_t \, ,
\label{eq:Langevin}
\end{equation}
where $W_t$ is a Wiener process, a scalar Brownian motion. The general non-linear functions $D^{(1)}(P \vert u)$ and $D^{(2)}(P \vert u)$ are the drift and the diffusion functions, which in case of the Langevin equation~(\ref{eq:Langevin}) are identical to the first and second order Kramers-Moyal (KM) coefficients. In general, the $j$-th order KM coefficients, $K^{(j)}(P \vert u)$, can be directly determined from given data $P$ for each wind speed $u$, using their definitions in terms of conditional incremental averaging, cf.~\cite{Friedrich2011,Tabar2019}, as
\begin{equation}
K^{(j)}(P \vert u) = \lim_{\Delta t \to 0} \frac{1}{\Delta t} \left\langle (P(t + \Delta t) -P(t))^j \vert_{P(t) = P, \, u(t) = u} \right\rangle \, .
\label{eq:KMcoef}
\end{equation}
The Langevin equation describes a continuous diffusion process where $K^{(j)}(P \vert u) = 0$ for $j \geq 3$ and $D^{(j)}(P \vert u) = K^{(j)}(P \vert u)$ for $j=1, 2$. Further details on methods of this estimation can be found in \cite{Friedrich2011, Rinn2016}.%
\footnote{KM coefficients of a Langevin process in $x(t)$ are defined for $j=1,2$ as $K^{(j)}(x,t) = D^{(j)}(x,t) = \frac{1}{j!}\lim_{\Delta t \to 0}\frac{1}{\Delta t}\left\langle\left( x(t + \Delta t) - x(t) \right)^{j} \vert_{x(t) = x}  \right\rangle$ in \cite{Friedrich2011, Rinn2016}. In order to stay consistent with the jump-diffusion process, our definition differs by a factor of $\frac{1}{j!}$, 
and $\mathrm{d}W_t = \int_t^{t+dt} \Gamma(\tau) \cdot \mathrm{d}\tau$ 
where $\left\langle \Gamma(t) \right\rangle = 0$ and $\left\langle \Gamma(t) \Gamma(t^{\prime}) \right\rangle = \delta(t-t^{\prime})$. The corresponding Fokker-Planck equation will be $\frac{\partial}{\partial t}p(x,t) = - \frac{\partial}{\partial x} \left[ D^{(1)}(x,t) \, p(x,t)\right] + \frac{1}{2}\frac{\partial^2}{\partial x^2} \left[ D^{(2)}(x,t) \, p(x,t)\right]$.} 
All the higher order KM coefficients vanish when the fourth order KM coefficient $K^{(4)}(P \vert u)$ is negligible according to the Pawula theorem \cite{Risken1996}. 
When the signal of a stochastic process has sharp changes, or discontinuities, at some instants, 
typically higher order Kramers-Moyal coefficients and especially $K^{(4)}(P \vert u)$ are not negligible anymore. In this case, an extension of the Langevin-type modeling with an additional jump noise is needed, see \cite{Tankov2003, Stanton1997, Johannes2004, Bandi2003, Anvari2016a, Tabar2019}. Such a jump-diffusion dynamics for a power conversion process is given by 

\begin{equation}
\mathrm{d}P(t, u) = D^{(1)}(P \vert u) \, \mathrm{d}t + \sqrt{D^{(2)}(P \vert u)} \, \mathrm{d}W_t + \xi \, \mathrm{d}J_t \, ,
\label{eq:jumpDiffusion}
\end{equation}
where again $W_t$ is a Wiener process, $D^{(1)}(P \vert u)$ and $D^{(2)}(P \vert u)$ are the drift and the diffusion functions. 
In the following we assume that $\xi \, \mathrm{d}J_t$ is a Poisson jump process. The coefficient $\xi$ is the jump size, which is assumed to be normally distributed, $\xi \sim N(0,\sigma_{\xi}^{2})$, with zero mean and variance $\sigma_{\xi}^{2}$. $\xi$ is also known as jump amplitude. $J_t$ is a Poisson jump process which is a zero-one jump process with jump rate $\lambda(P \vert u)$ \cite{Hanson2007, Tabar2019}. The drift and diffusion coefficients and the jump rate are now related to the KM coefficients $K^{(j)}(P \vert u)$ in the following way \cite{Anvari2016a}:
\begin{equation}
D_\mathrm{j}^{(1)}(P \vert u) = K^{(1)}(P \vert u),
\label{eq:drift_jump}
\end{equation}
\vspace{-18pt}
\begin{equation}
D_\mathrm{j}^{(2)}(P \vert u) + \lambda(P \vert u)\langle \xi^2 \rangle = K^{(2)}(P \vert u),
\label{eq:diffusion_jump}
\end{equation}
\vspace{-18pt}
\begin{equation}
\lambda(P \vert u)\langle \xi^j \rangle =  K^{(j)}(P \vert u) \hspace{12pt} \textrm{for} \hspace{6pt} j > 2.
\label{eq:jumprate}
\end{equation}
Since $\xi$ has zero mean, its second order moment is $\langle \xi^2 \rangle = \sigma_{\xi}^{2}$. From Eq.~(\ref{eq:drift_jump}), it can be seen that the estimation of the drift coefficient is the same for the diffusion process, which obeys the Langevin equation, and the jump-diffusion process. Here we go into more details of the noise part and do not assume anymore a vanishing $K^{(4)} =0$.

Jump amplitude $\sigma_{\xi}^{2}$ and jump rate $\lambda$ can be estimated by using Eq.~(\ref{eq:jumprate}) with $j = 4$ and $6$ and Wick's theorem \cite{Isserlis1916, Wick1950} for Gaussian random variables, i.e., $\langle \xi^{2n} \rangle = \frac{(2n)!}{2^n n!} \langle \xi^2 \rangle^{n}$, 
\begin{equation}
\sigma_{\xi}^{2}(P \vert u) = \frac{K^{(6)}(P \vert u)}{5 K^{(4)}(P \vert u)} ,
\label{eq:jumpamp}
\end{equation}
\vspace{-6pt}
\begin{equation}
\lambda(P \vert u) = \frac{K^{(4)}(P \vert u)}{3 \sigma_{\xi}^{4}(P \vert u)} .
\label{eq:jumprate2}
\end{equation}

\subsection{Results}
\label{sec:results}

\subsubsection{Results for Electrical Power Output}
\label{sec:results-power}

\begin{figure*}[t]
\centering
\includegraphics[width = \textwidth]{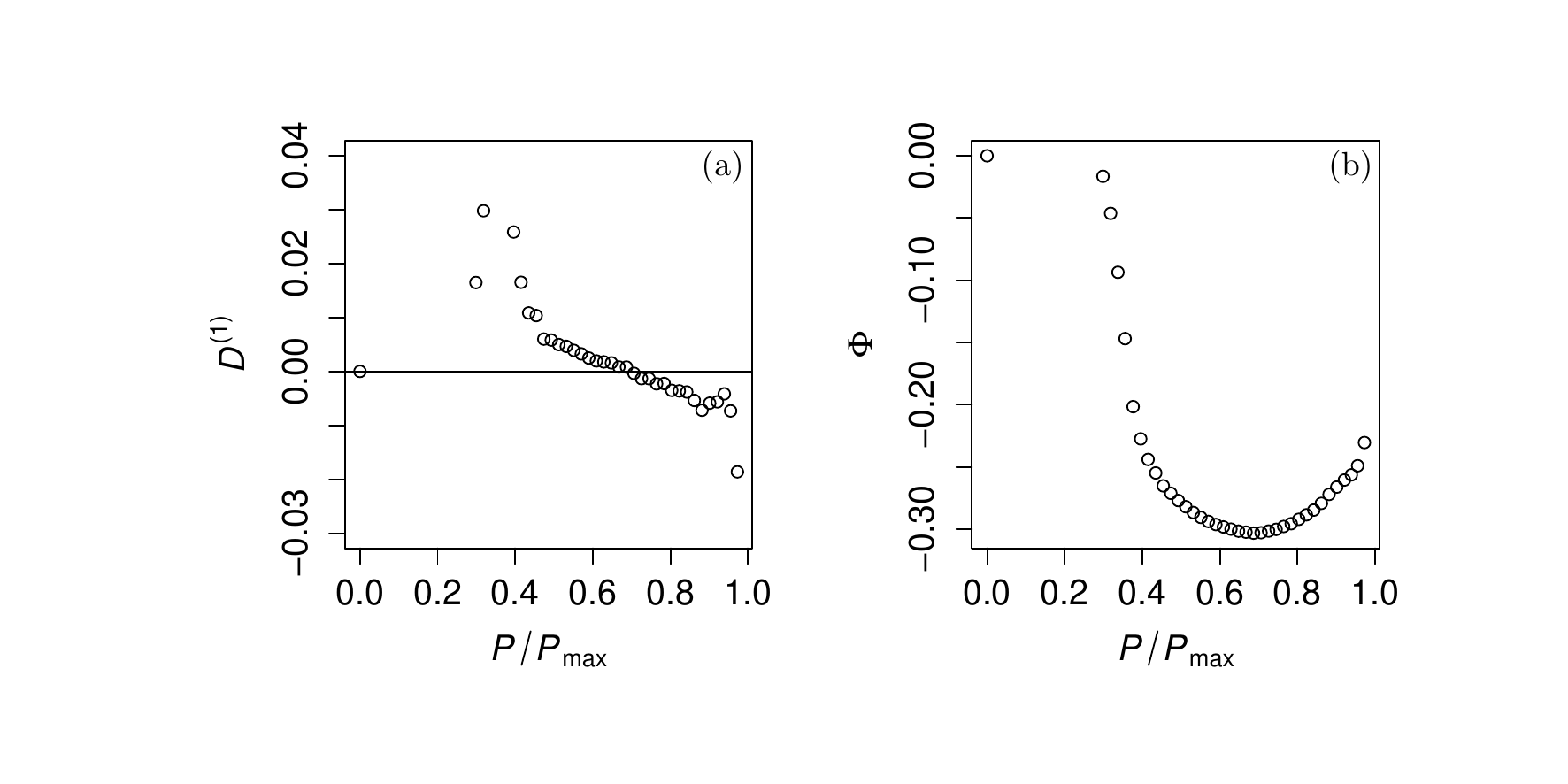}
\caption{Drift coefficient $D^{1}(P \vert u)$ (a) and corresponding potential $\Phi$ (b) for the wind speed of $u = 0.41\,u_\mathrm{max}$. Zero-crossings of the drift coefficient $D^{1}(P \vert u) = 0$ or local minima of drift potential $\Phi$ are stable fixed points which describe the equilibrium dynamics.}
\label{fig:drift}
\end{figure*}

First, we analyze the relation between wind speed and power. For chosen fixed wind speed values with bin sizes of $0.5\,{\rm ms^{-1}}$, the KM coefficients $K^{(j)}(P \vert u)$ are determined with the assumption of stationarity within the corresponding wind speed bin. Firstly, the drift coefficients are determined. The zero-crossings of the drift coefficient, $D^{(1)}(P \vert u) = 0$, correspond to the stable fixed points or equilibria of each wind speed bin if the slope of $D^{(1)}$ is negative \cite{Anahua2008, Gottschall2007}. Zero-crossings with positive slope are unstable fixed points. Alternatively, this can be expressed by a drift potential, which is defined as $\Phi = -\int_P D^{(1)}(P \vert u) \; \mathrm{d}P$. The zero-crossings with negative slope of the drift correspond correspond to minima of the drift potential. An example of a drift coefficient and corresponding potential for the wind speed of $u = 0.41\,u_\mathrm{max}$ is shown in Fig.~\ref{fig:drift}.

\begin{figure*}[t]
\centering
\includegraphics[width = \textwidth]{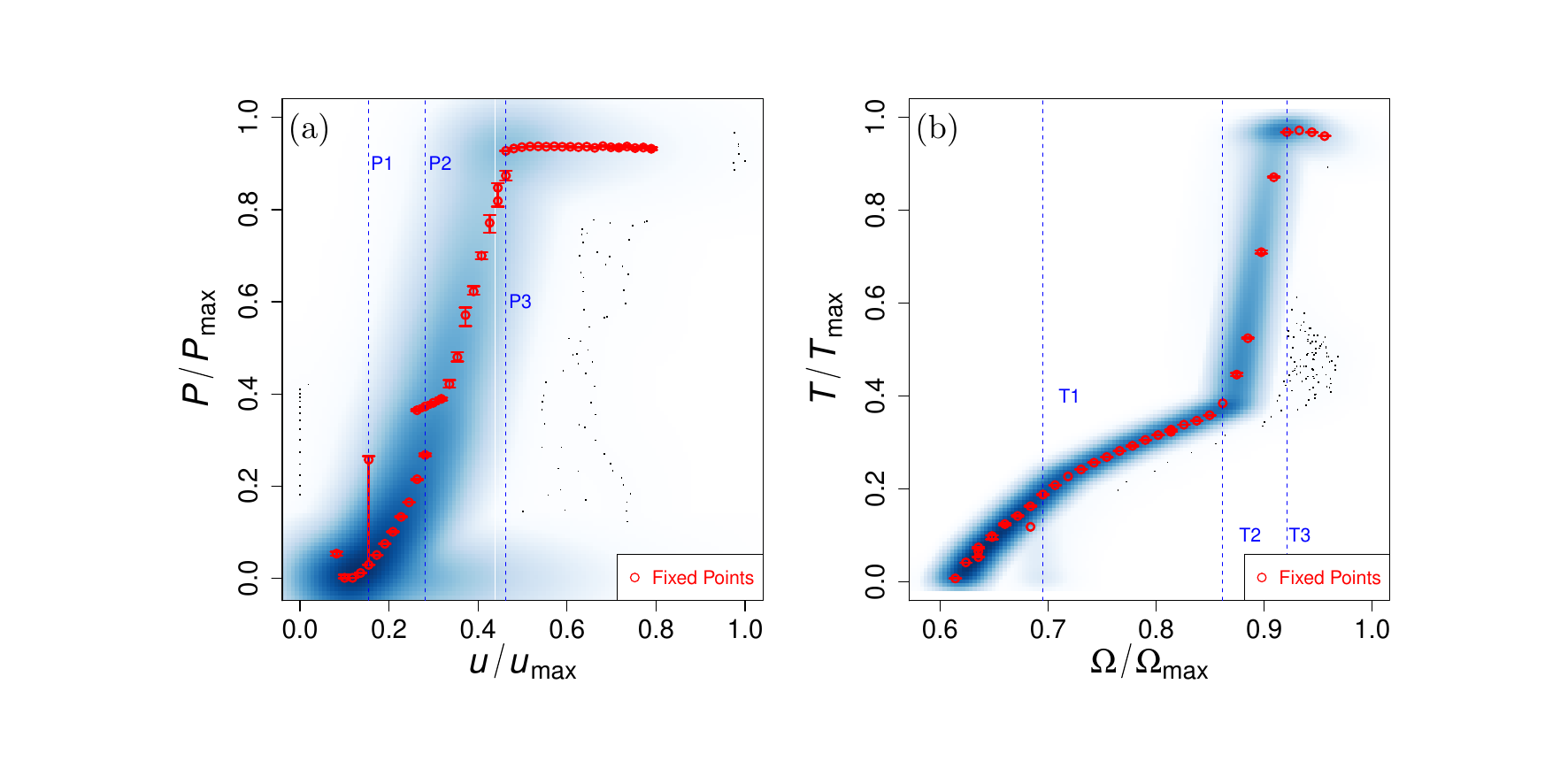}
\caption{Characteristic power curve, (a), determined from the zero crossings of the drift coefficients at each wind speed bin and characteristic torque curve, (b), at each rotational speed rpm bin, presented in red open circles. They are also called Langevin Power Curve (LPC) and Langevin Torque Curve (LTC), respectively. The blue background shows the density scatter plot of the measurement data and darker regions indicate more data points are available. The black dots are the outliers of the density scatter plot. Three distinct states (P1, P2 and P3) for LPC and (T1, T2 and T3) for LTC which separate the operational regions are marked by blue dotted lines.}
\label{fig:LPC}
\end{figure*}

For each wind speed bin, there can be single or multiple fixed points. With these stable fixed points, we can reconstruct the characteristic power curve, which we call Langevin Power Curve (LPC) \cite{Milan2010, Waechter2011}, as shown in Fig.~\ref{fig:LPC}~(a). These stable fixed points can already be used for a definition of different operational states of the wind turbine. In our case, we mark three distinct states (P1, P2 and P3) which separate the operational regions by blue dotted lines.
Multiple fixed points are found to be at these states. Near operation point P2 in Fig.~\ref{fig:LPC}~(a), we observe the shifting of fixed points in a discontinuous way. Such details cannot be detected by the standard averaging procedure of power curve defined by \cite{IEC61400}.

The fixed point analysis and characterization of power output of a wind turbine by Langevin equation (\ref{eq:Langevin}) or diffusion process have been studied by \cite{Anahua2008, Gottschall2008, Milan2014}. In their works, they extensively focused on the drift coefficient. Higher order KM coefficients were not considered. In our work here, we focus on the noisy part and evaluate the higher order KM coefficients.

As explained in Sec.~\ref{sec:sde}, first the fourth-order KM coefficient $K^{(4)}(P \vert u)$ is estimated to see if jump noise matters. An example of $K^{(4)}(P \vert u)$ and the jump amplitude $\sigma_\xi^2 (P \vert u)$ for the wind speed of $u = 0.41\,u_\mathrm{max}$ with their medians (solid black lines) is shown in Fig.~\ref{fig:K4jump}. The fixed point for this wind speed bin is located at $P = 0.7\,P_\mathrm{max}$, see Fig.~\ref{fig:drift}. 
Statistically, we can obtain more accurate results near the fixed point due to the better coverage of data, whereas for regions with less data (farther away from the fixed point) the results become more noisy and outliers are seen. A robust method to estimate is to use the medians instead of the means \cite{Huber2011}. Some examples are shown in Appendix~\ref{app:median}.

\begin{figure*}[t]
\centering
\includegraphics[width = \textwidth]{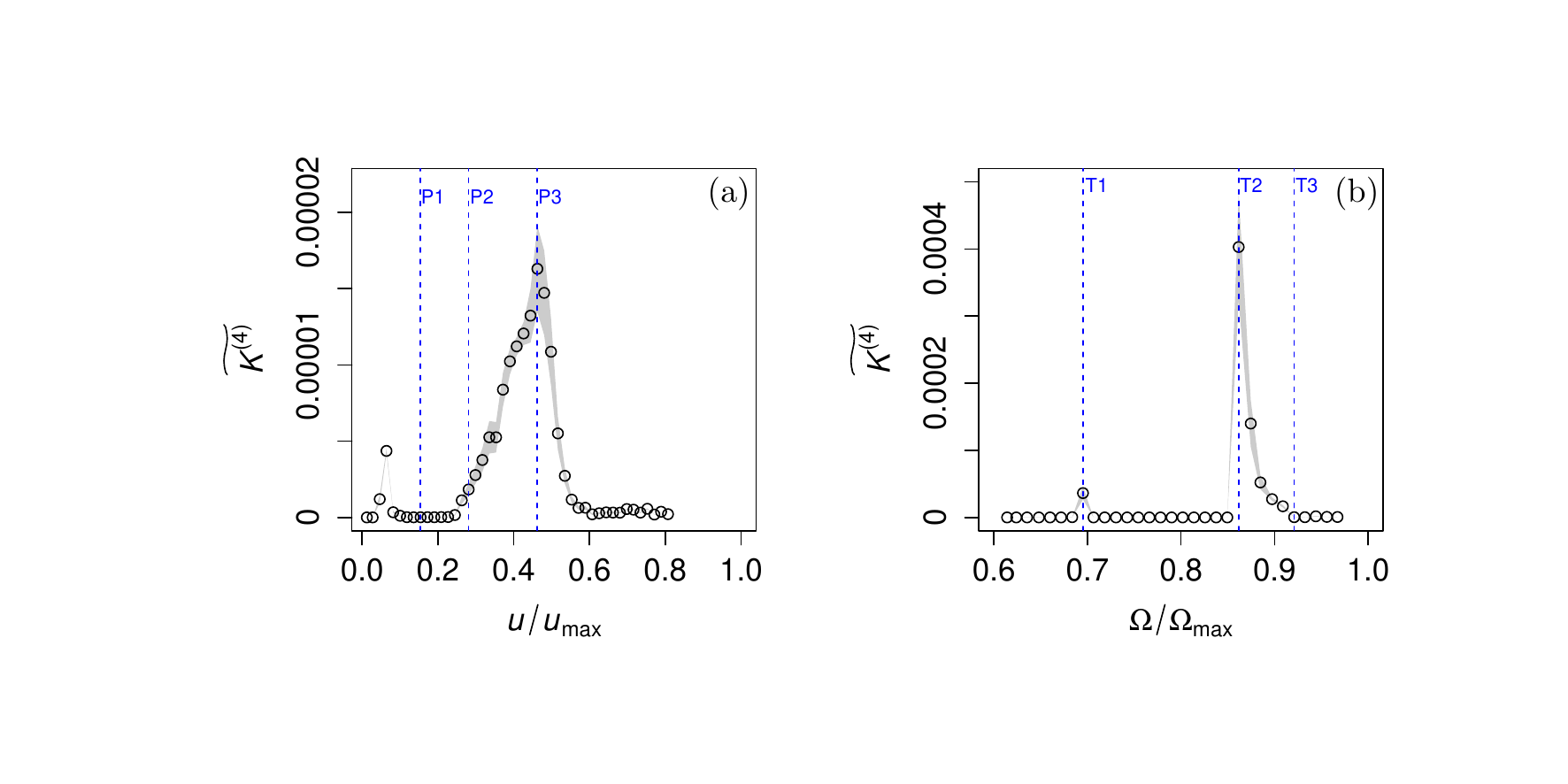}
\caption{The median of the fourth-order KM coefficient, $\widetilde{K^{(4)}}(u)$ over the power bins for each wind speeds bin (a) and over the torque bins for each rpm bin (b). There is an increase of $\widetilde{K^{(4)}}(u)$ near the transition to the rated region. Statistical uncertainties are shown as gray-shaded background. Blue dotted lines are the distinct states observed from the fixed point analysis, see Fig.~\ref{fig:LPC}.}
\label{fig:K4}
\end{figure*}

\begin{figure*}[t]
\centering
\includegraphics[width = \textwidth]{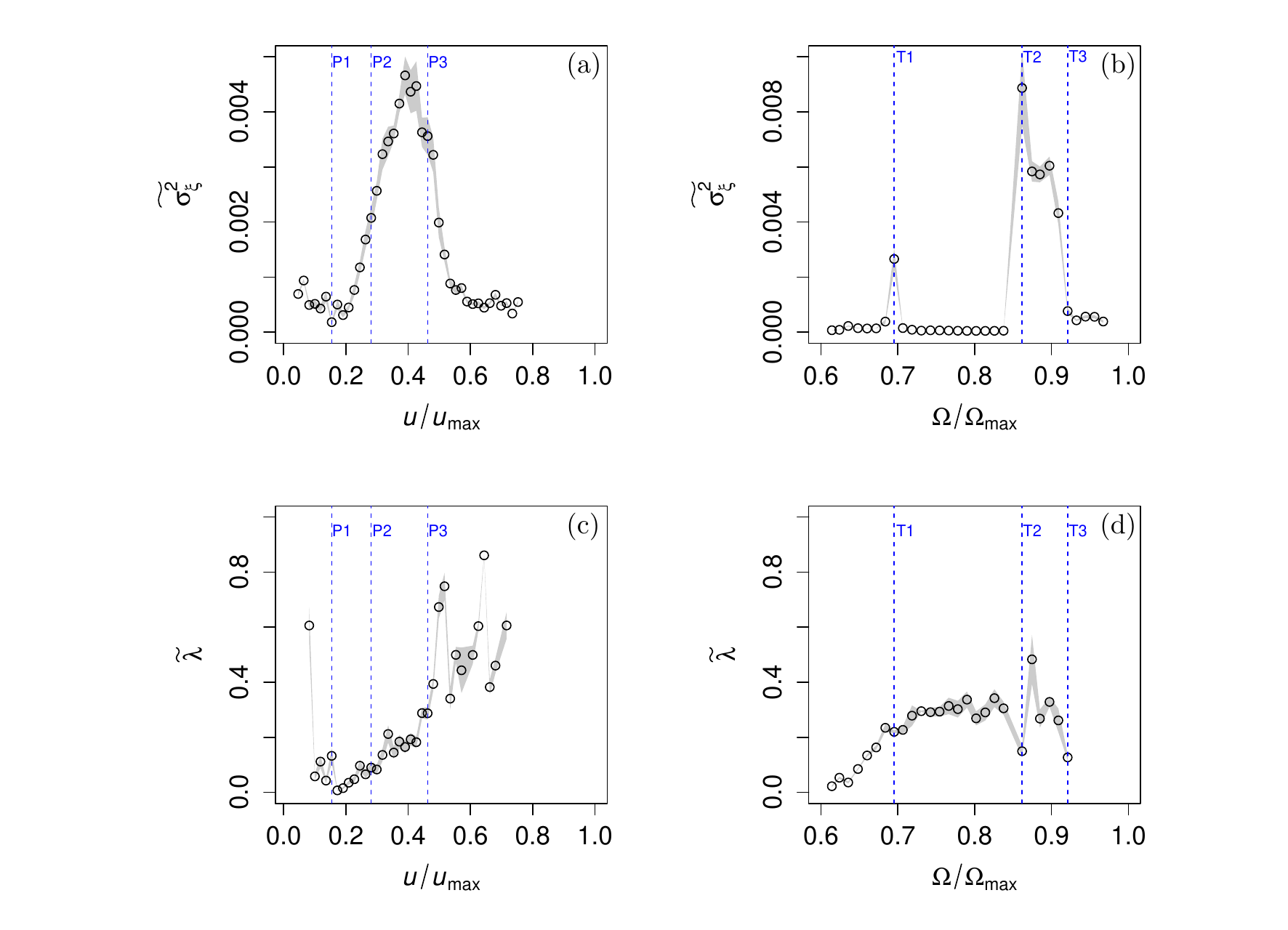}
\caption{The median of jump amplitude, $\widetilde{\sigma_\xi^2}$, over the power bins for each wind speed bin (a) and over the torque bins for each rpm bin (b). There is a significant increase in jump amplitude near the transition to the rated region. The median of jump rate, $\widetilde{\lambda}$, over the power bins for each wind speed bin (c) and over the torque bins for each rpm bin (d). There is a significant increase in jump rate after the transition to the rated region and a small increase near the cut-in region for power analysis. The jump rate for torque analysis is similar in between all three 
distinct states shown by blue dotted lines as in Fig.~\ref{fig:LPC}. Statistical uncertainties are shown as gray-shaded background.}
\label{fig:jump}
\end{figure*}

In the following, we investigate details of the median values. Thus, we simplify the process to those with constant parameters $\sigma_\xi^2$ and $\lambda$ for the jump process. The $P$-dependence of these parameters can also be studied, which we do not do here to keep the discussion simpler, see Fig.~\ref{fig:K4jump}. Fig.~\ref{fig:K4}~(a) shows that there is an increase of $\widetilde{K^{(4)}}(P \vert u)$ near the state P3, the transition point to the rated power. This behavior of $\widetilde{K^{(4)}}$ shows that not only diffusive noise is present, thus we proceed to analyze also the higher order KM-coefficients from which we can determine the jump amplitude $\widetilde{\sigma_\xi^2}$ and jump rate $\widetilde{\lambda}$ for each wind speed as shown in Fig.~\ref{fig:jump}~(a) and (c). The jump amplitude $\sigma_\xi^2$ is highest between the states P2 and P3 which is just below the transition to the rated power region where the switching of the control strategy play a major role. The jump rate $\widetilde{\lambda}$ is highest in the region of rated power. 

\begin{figure*}[t]
\centering
\includegraphics[width = \textwidth]{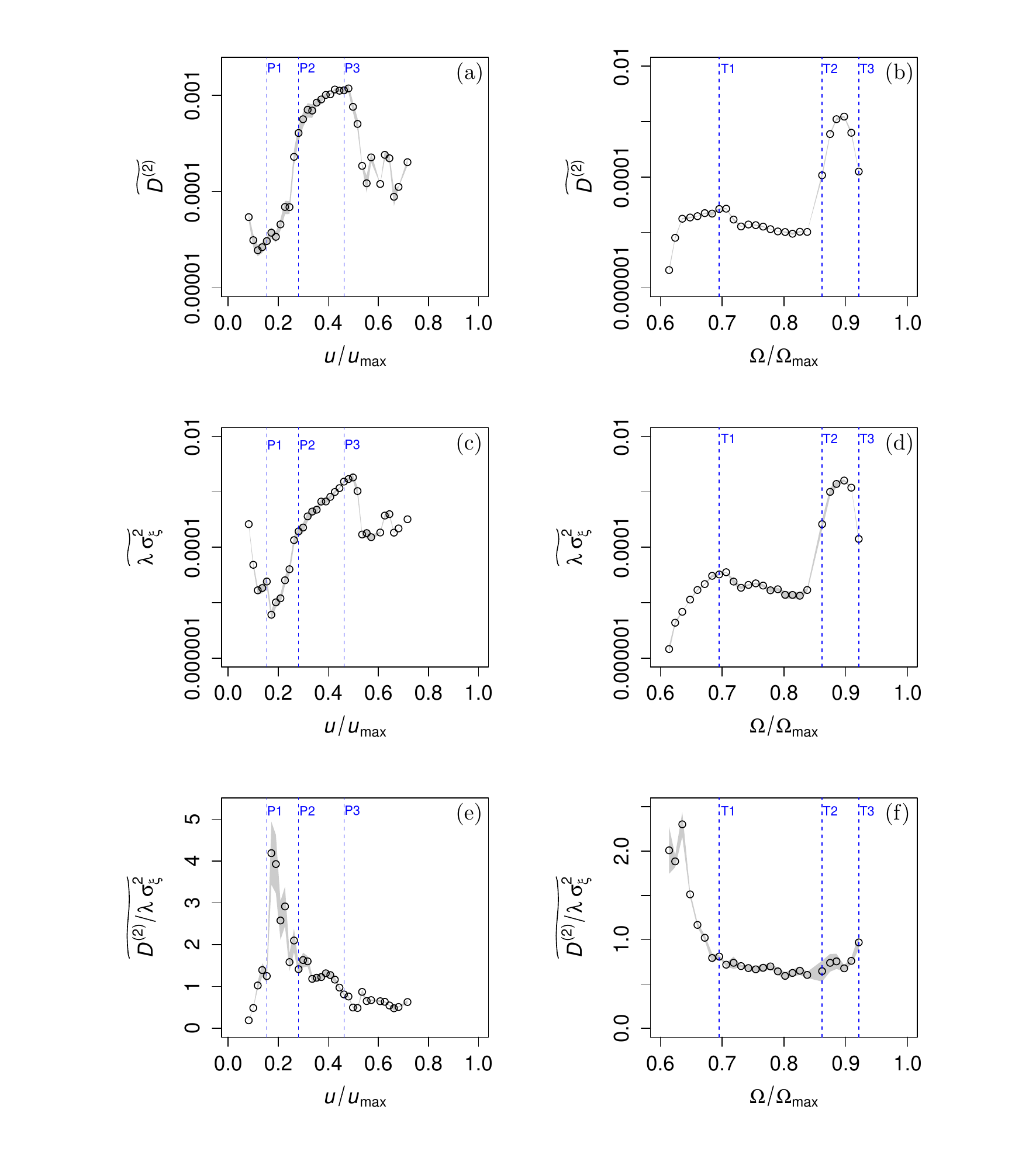}
\caption{The median of diffusion coefficient, $\widetilde{D^{(2)}}$, (a), overall jump contribution, $\widetilde{\lambda \sigma_\xi^2}$, (c), and the diffusion to jump ratio, $\widetilde{\frac{D^{(2)}}{\lambda \sigma_\xi^2}}$, (e), over the power bins for each wind speed bin, and respective quantities over the torque bins for each rpm bin in (b), (d), and (f). Sub-figures (a-d) are plotted in semi-logarithmic scale for better visualization. Statistical uncertainties are shown as gray-shaded background. Blue dotted lines are the distinct states observed from fixed point analysis, see Fig.~\ref{fig:LPC}.}
\label{fig:diffjump}
\end{figure*}

In order to quantify the overall jump contribution, we determine the product $\widetilde{\lambda \sigma_\xi^2}$ as shown in Fig.~\ref{fig:diffjump}~(c). 
Again we see that the jump contribution $\widetilde{\lambda \sigma_\xi^2}$ is highest around the 
state P3. 
Moreover, we also determine the median of the diffusion to jump ratio $\widetilde{\frac{D^{(2)}}{\lambda \sigma_\xi^2}}$ over the power bins for each wind speed bin obtained from the resolved $\frac{D^{(2)}}{\lambda \sigma_\xi^2}$ values as shown in Fig.~\ref{fig:diffjump}~(e) to find out whether diffusive or jump noise is dominating. If the ratio is large, there is more diffusive noise and vice versa. The jump noise is dominant in the rated power region after the state P3. The diffusive noise is dominant between the states P1 and P2, which also coincides with the lowest jump contribution $\lambda \sigma_\xi^2$, Fig.~\ref{fig:diffjump}~(c). 

The analysis in this subsection shows two important points. First, a jump process is present and should be included in an advanced stochastic description or, respectively, model. Second, below rated power, first a diffusive stochastic behavior is dominating, while jumpy noise becomes important for the considered wind turbine close to the transition to rated power.

\subsubsection{Results for Generator Torque}
\label{sec:results-torque}

Apart from the dynamical dependence of power on the wind speed, another 
central characteristic of the wind turbine is the generator torque $T$ vs. rotational speed or rpm $\Omega$ \cite{Lind2014}. The generator torque was calculated from power and rpm data according to Eq.~(\ref{eq:torq}), whereas the rpm was measured independently. 
A similar stochastic approach like for the power and wind speed is applied to the torque and rpm dynamics $T(t,\Omega)$, next. Based on the drift coefficient, a characteristic curve like the Langevin Power Curve 
is calculated as shown in Fig.~\ref{fig:LPC}~(b). We refer to it as Langevin Torque Curve (LTC). We again mark three distinct 
states (T1, T2 and T3) with blue dotted lines which can be deduced from the fixed point analysis. 

Next, we also determine the jump contribution $\lambda \sigma_\xi^2$ and the diffusion to jump ratio $\frac{D^{(2)}}{\lambda \sigma_\xi^2}$ for each rpm bin, see Fig. \ref{fig:diffjump}~(d) and (f). 
In Fig.~\ref{fig:diffjump}~(d), we see that the median of the jump contribution $\widetilde{\lambda\sigma_\xi^2}$ becomes significant in the region between T2 and T3, cf.\ Fig.~\ref{fig:LPC}~(b), at relatively high rpm values. Outside this regime, jump noise does not dominantly contribute to the dynamics. On the other hand, the median of the diffusion to jump ratio $\widetilde{\frac{D^{(2)}}{\lambda \sigma_\xi^2}}$ takes its largest values at low values of rotational speed rpm. At the rotational speeds rpm below the state T1, diffusive behavior is much more dominant. 

The conclusions from the previous subsection \ref{sec:results-power} apply also to the torque analysis in an analogous way.
Similar to the power characteristic, a jump process is also present for the torque characteristic, which could be expected as we compute the torque from power and rotational speed. Diffusive stochastic behavior is dominating for low rpm, but in the region from T1 on ($\Omega/\Omega_\mathrm{max}\gtrsim 0.7$) diffusive and jumpy behavior seem to be more balanced for the torque case. 

\section{Conclusion and Outlook}

In our work, we investigate the contribution of the higher-order KM coefficients to the stochastic conversion dynamics of a wind turbine. As described in Sec.~\ref{sec:sde}, these higher-order coefficients allow to quantify the contributions of diffusive behavior and jump noise, and indicate that discontinuities in the trajectory of the measurement data are due to the stochastic jump noise. The main results are that we can quantify with our proposed method how the amplitudes and the ratio of the two noise contributions change in different operating ranges of a wind turbine. The region below rated power seems to provide the highest values of the amplitudes ($D^{(2)}$ and $\sigma_\xi^2$). Sometimes the maximal values are found for the transition states defined by the fixed point characteristics. The ratio between the contributions of the diffusive and jumpy noise shows that at low wind speed and low power a diffusive noise is dominating whereas for higher power more jump noise is present, with some detailed differences for power and torque. All this indicates that it is the interplay between the stochastic driving wind speed and the reacting control system that determines the noise contribution. In particular, the jump contribution is closely linked to the control system as one can see, to our interpretation, in the rapid changes of $\sigma_\xi^2$ in Fig.~\ref{fig:jump}~(b). Interestingly, this is more prominent in the torque signal than in the power signal. It is well known that the control system is not operating directly with the wind signal but with toque $T$ and the rotational speed $\Omega$. 

Near the states T1, T2 and T3,  there are three distinct operational rotational speeds $\Omega$ which the control system prefers to approach. It is a common control strategy to avoid certain resonance frequencies of the structure in order to mitigate excessive loads. We show this by evaluating the drift potential $\Phi(\Omega)$ of the rotational speed. The minima of this potential correspond to the preferred rotational speeds, see Appendix~\ref{app:rpmpot}. Moreover, looking at Fig.~\ref{fig:LPC}~(b), between the states T2 and T3, there is a steep gradient which enforces a large change in generator torque $T$ at only a small regime in rotational speed $\Omega$. In this range, we also observe that there is a huge increase in both diffusive and jump noise by more than two orders of magnitude, see Fig.~\ref{fig:diffjump}~(b) and (d).

So far we used the stochastic methods to characterize the dynamics of the wind energy conversion process. It goes without saying that the characterization can be used to compare quantitatively different turbines. Potential failures in the control system should be detectable by comparison of the different stochastic terms. One may see how with time some noise contribution changes as the system gets old, or one may show how different wind turbines or different contril strategies perform differently in a dynamic sense. Together with detailed knowledge of a specific turbine this should also be useful for monitoring, e.g., performance or structural health. 

At last we would like to point out that besides this characterization the stochastic methods presented here also deliver the explicit form of the stochastic differential equations. Thus, it is also possible to use our results as very efficient dynamics models for power and torque. Long time simulations can be done easily. Such models are of interest for the simulation of the contribution of wind energy to the power grid and for the simulation of loads.

\begin{acknowledgments}
The authors would like to thank Vlaho Petrovi\'c and Christian Philipp for helpful discussions. We acknowledge financial support by the Federal Ministry for Economic Affairs and Climate Action of Germany in the framework of the projects ``WEA-Doktor'' (reference 0324263A) and ``WiSAbigdata'' (reference 03EE3016A).\\
\end{acknowledgments}

\appendix

\section{Median as a Robust Estimator}
\label{app:median}
\begin{figure*}[t]
\centering
\includegraphics[width = \textwidth]{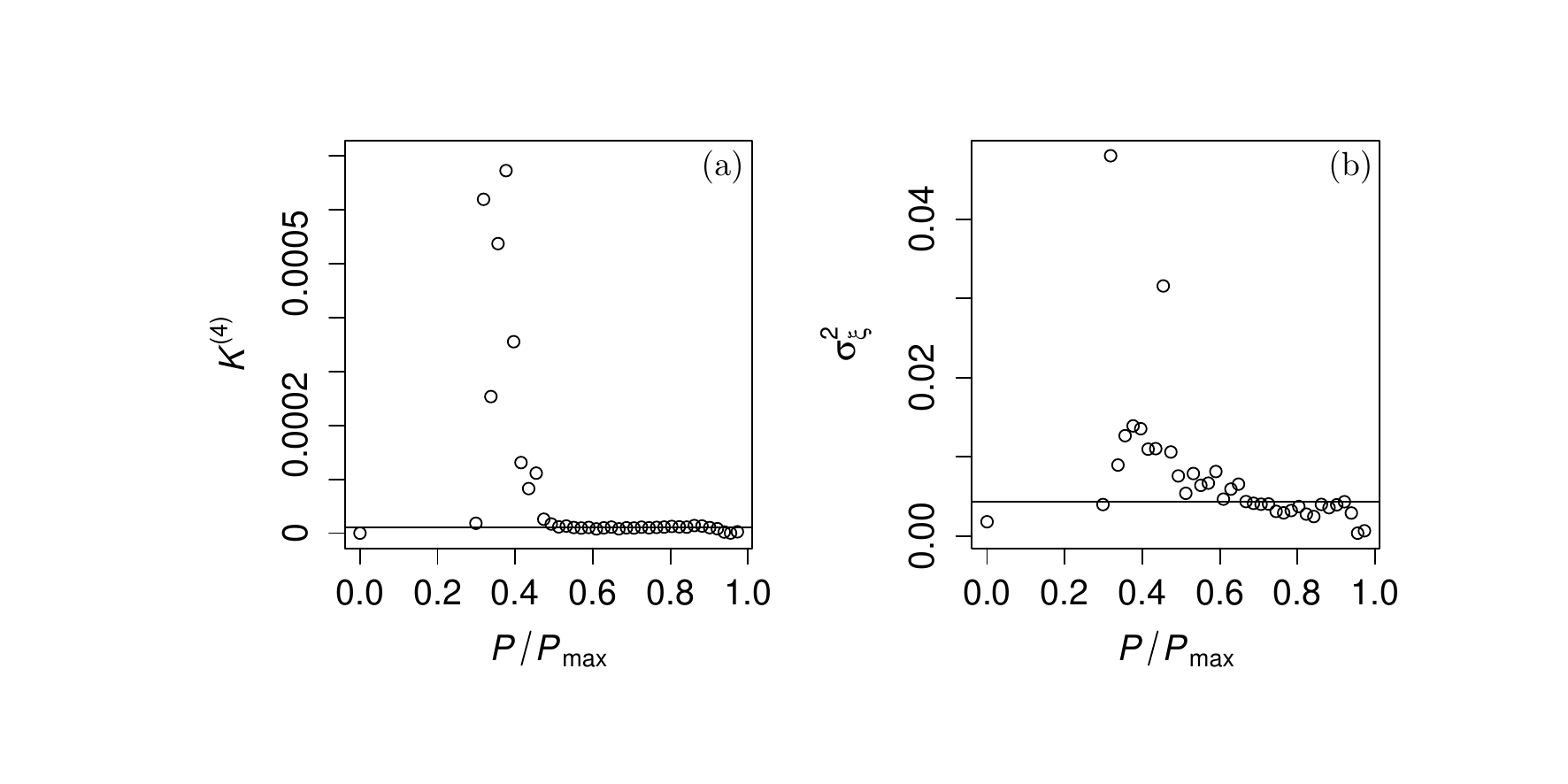}
\caption{Fourth-order KM coefficient $K^{(4)}(P \vert u)$ (a) and the jump amplitude $\sigma_\xi^2 (P \vert u)$ (b) for the wind speed of $u = 0.41\,u_\mathrm{max}$. (For the corresponding drift term see Fig.~\ref{fig:drift}). The solid black lines are their respective medians $\widetilde{K^{(4)}}(u)$ and $\widetilde{\sigma_\xi^2} (u)$. The fixed point for this wind speed bin is $P = 0.7\,P_\mathrm{max}$. Statistically, more accurate results can be obtained near the fixed point due to the better coverage of data. By using the median, our results are more robust to outliers far away from the fixed points.}
\label{fig:K4jump}
\end{figure*}

Statistically, we can obtain more accurate results near the fixed point due to the better coverage of data, whereas for regions with less data (farther away from the fixed point) the results become more noisy and outliers are seen. A robust method to estimate the typical value of $K^{(4)}(P \vert u)$ and  $\sigma_\xi^2 (P \vert u)$, as examples, is to use the medians $\widetilde{K^{(4)}}(u)$ and $\widetilde{\sigma_\xi^2}(u)$ as shown by solid lines in Fig.~\ref{fig:K4jump}.

\section{Drift Potential of Rotational Speed}
\label{app:rpmpot}

\begin{figure*}[t]
\centering
\includegraphics[width = \textwidth]{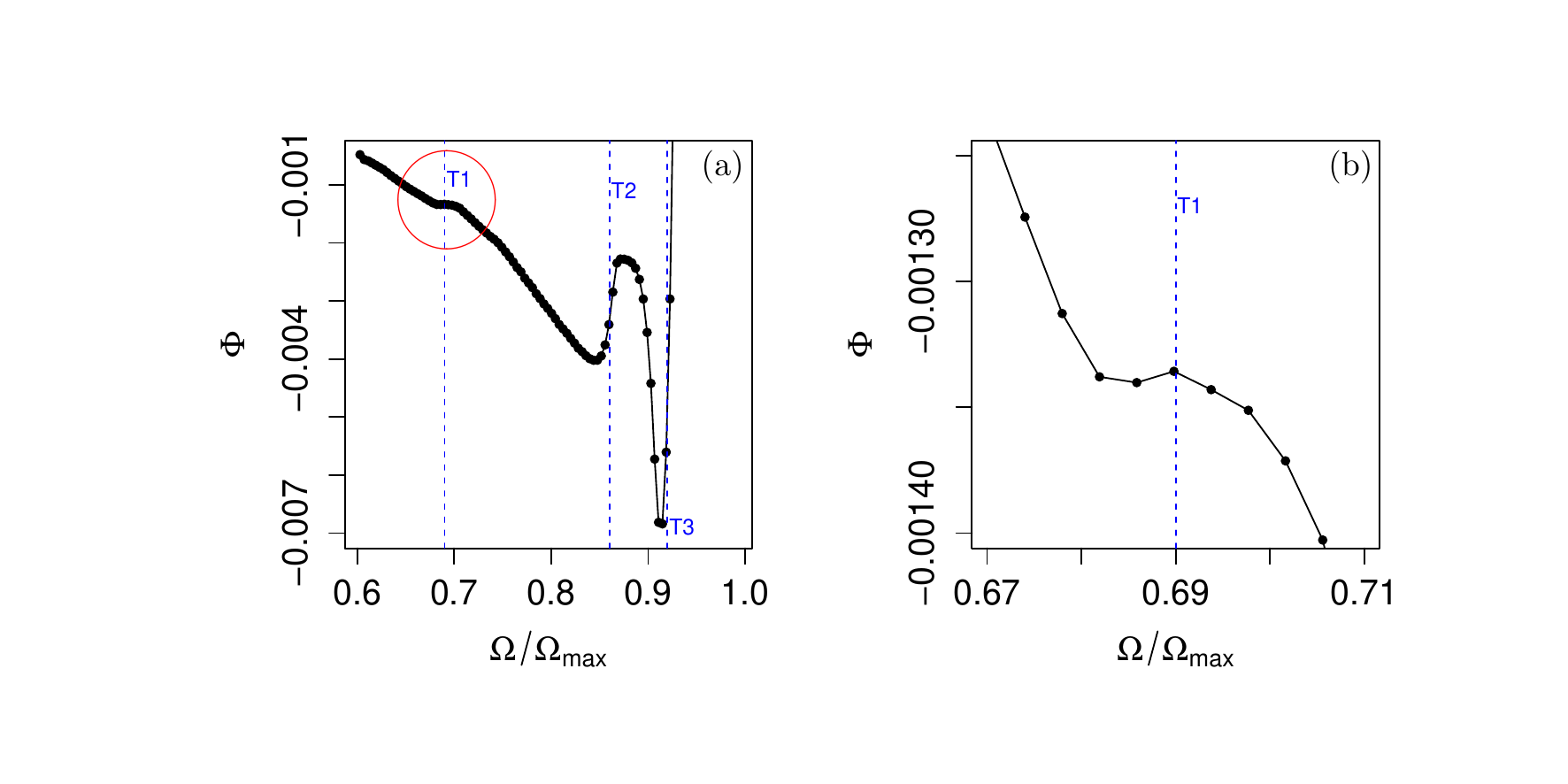}
\caption{Drift potential $\Phi(\Omega)$ determined in the range of $0.6~\Omega_{\rm max}$ and $\Omega_{\rm max}$, (a). Minima of the drift potential corresponds to the stable fixed points or equilibria. Here we can observe three minima around the three 
states T1, T2 and T3 as shown in Fig. \ref{fig:LPC}. The minimum around the state T1 which is presented with the red circle is elaborated in (b).}
\label{fig:rpmpot}
\end{figure*}

We analyzed all the data of rotational speed $\Omega$ in the range of $0.6~\Omega_{\rm max}$ and $\Omega_{\rm max}$ without any conditioning or binning on other variables. We evaluated the drift coefficient $D^{(1)}(\Omega)$ and then determined the drift potential which is  $\Phi(\Omega) = -\int_\Omega D^{(1)}(\Omega) \; \mathrm{d}\Omega$ which is plotted in Fig.~\ref{fig:rpmpot}. Minima of the drift potential corresponds to the stable fixed points or equilibria. Here we can observe three minima around the three 
states T1, T2 and T3 which the control system prefers to approach. It is a common control strategy to avoid certain resonance frequencies of the structure in order to mitigate excessive loads as shown in Fig.~\ref{fig:LPC}.

In Fig.~\ref{fig:rpmpot}~(b), we can clearly observe a minimum around the state T1. From our results in Fig.~\ref{fig:K4}, \ref{fig:jump} and \ref{fig:diffjump}, there is also a slight increase in noises around this state. This indicates that the control system of the wind turbine starts switching the strategies at this state around T1. As a remark, we calculated the deterministic potential only which reflects the mechanical and control mechanism of the wind turbine.

\bibliography{ref.bib}

\end{document}